# Extraordinarily bound quasi-one-dimensional trions in two-dimensional phosphorene atomic semiconductors

Shuang Zhang,[1,†] Renjing Xu,[1,†] Fan Wang,[2] Jiong Yang,[1] Zhu Wang,[3] Jiajie Pei,[1] Ye Win Myint,[1] Bobin Xing,[1] Zongfu Yu,[3] Lan Fu,[2] Qinghua Qin,[1] and Yuerui Lu[1*]

[1]Research School of Engineering, College of Engineering and Computer Science, the Australian National University, Canberra, ACT, 0200, Australia

[2]Department of Electronic Materials Engineering, Research School of Physics and Engineering, the Australian National University, Canberra, ACT, 0200, Australia

[3]Department of Electrical and Computer Engineering, University of Wisconsin, Madison, Wisconsin 53706, USA

[†] These authors contributed equally to this work

[*] To whom correspondence should be addressed: Yuerui Lu (yuerui.lu@anu.edu.au)

The anisotropic nature of the new two-dimensional (2D) material phosphorene[1-9], in contrast to other 2D materials such as graphene[10] and transition metal dichalcogenide (TMD) semiconductors[11-13], allows excitons to be confined in a quasi-one-dimensional (1D) space predicted in theory[8,9], leading to remarkable phenomena arising from the reduced dimensionality and screening. Here, we report a trion (charged exciton) binding energy of 190 meV in few-layer phosphorene at room temperature, which is nearly one to two orders of magnitude larger than those in 2D TMD semiconductors[12,13] (20–30 meV) and quasi-2D quantum wells[14,15] (~1–5 meV). Such a large binding energy has only been observed in truly 1D materials such as carbon nanotubes[16,17], whose optoelectronic applications have been severely hurdled by their intrinsically small optical cross-sections. Phosphorene offers an elegant way to overcome this hurdle by enabling quasi-1D excitonic and trionic behaviors in a large 2D area, allowing optoelectronic integration. We experimentally validated the quasi-1D nature of excitonic and trionic dynamics in

**phospherene by demonstrating completely linearly polarized light emission from excitons and trions. The implications of the extraordinarily large trion binding energy in a higher-than-one-dimensional material are far-reaching. It provides a room-temperature 2D platform to observe the fundamental many-body interactions in the quasi-1D region. The strong photoluminescence emission in phosphorene has been electrically tuned over a large spectral range at room temperature, which opens a new route for tunable light sources[18,19].**

A neutral exciton is a bound quasi-particle state between one electron and one hole through a Coulomb interaction, similar to a neutral hydrogen atom. A trion is a charged exciton composed of two electrons and one hole (or two holes and one electron), analogous to $H^-$ (or $H_2^+$)[20]. Trions have been of considerable interest for the fundamental studies of many-body interactions, such as carrier multiplication and Wigner crystallization[21]. In contrast to the exciton, a trion has an extra charge with nonzero spin, which can be used for spin manipulation[22,23]. More importantly, the density of trions can be electrically tuned by the gate voltage, enabling remarkable optoelectronic applications[18-20,24,25]. For these purposes, a large trion binding energy is critical in order to overcome the room-temperature thermal fluctuations as well as to widen the spectral tuning range. The dimensional confinement is the dominating factor that determines the binding energy of trions. In quasi-2D quantum wells, the trion binding energy is only 1–5 meV, and trions are highly unstable, except at cryogenic temperatures[16,17]. Recently, Shan[14] and Xu[15] made an important breakthrough, showing that truly 2D atomic TMD semiconductors have trion binding energies up to 20–30 meV, which is still barely resolvable at room temperature compared with their emission bandwidth. On the other hand, trions in the 1D space, such as carbon nanotubes, exhibit remarkably higher binding energies in the range of 100–200 meV owing to the stronger Coulomb interaction with the reduced dimensionality and screening[16,17]. The complete separation of the exciton and trion emission peaks was observed at room temperature[16,17]. However, the application of 1D carbon nanotubes for practical optoelectronic devices is intrinsically limited by their small cross-sections. The overall optical responses of such 1D lines are extremely weak. The diverse distribution of the chirality in carbon nanotubes also makes it impossible to assemble a large-size film with uniform optoelectronic responses. While the reduced dimensionality leads to far more attractive exciton and trion properties, the trade-off between the cross-section and the dimensional confinement has hindered the development of useful excitonic optoelectronic devices.

Here, we show that phosphorene presents an intriguing platform to overcome the aforementioned trade-off. We observed quasi-1D trions with ultra-high binding energies up to ~190 meV in 2D phosphorene atomic semiconductors. Using back-gated metal-oxide-semiconductor (MOS) devices, we demonstrated the reversible electrostatic tunability of the exciton charging effects between trions and excitons in few-layer phosphorene. The measured ultra-high trion binding energies (~190 meV), comparable to those in truly 1D semiconductors[16,17], are due to the formation of quasi-1D trions and excitons in 2D phosphorene. The quasi-1D excitons and trions in phosphorene were demonstrated by our measured linearly polarized PL emission from the excitons and trions. In this regard, phosphorene is equivalent to a system that is made of a bundle of identical 1D materials. In addition, the measured layer-dependent trion binding energies in few-layer phosphorene agree well with our theoretical calculations. Our results open exciting avenues for optoelectronic applications, including tunable light sources[18,19], spin manipulation devices[22,23], and quantum logical systems[24,25].

This new type of material, few-layer phosphorene, is unstable and does not survive well in many standard nanofabrication processes. To overcome the challenge of the instability, we designed special fabrication and characterization techniques. We used mechanical exfoliation to drily transfer[26] a phosphorene flake onto a SiO$_2$/Si substrate (275 nm thermal oxide on n$^+$-doped silicon). The phosphorene was placed near a gold electrode that was pre-patterned on the substrate. Another thick graphite flake was similarly transferred to electrically bridge the phosphorene flake and the gold electrode, forming a MOS device (Figure 1). This fabrication procedure kept the phosphorenes free from chemical contaminations by minimizing the post-processes after the phosphorene flake was transferred. In the measurement, the gold electrode is grounded, and the n$^+$-doped Si substrate functions as a back gate providing uniform electrostatic doping in the phosphorene (Figure 1b).

Another unique challenge in working with unstable phosphorene is the identification of the layer number. Atomic force microscopy (AFM) is typically unreliable for the identification of very-few-layer phosphorene (one or two layers) because of its slow scanning speed and the potential contact contamination. Here, we propose and implement a rapid, noninvasive, and highly accurate approach to determine the layer number using optical interferometry. Specifically, we measured the optical path length (OPL) of the light reflected from phosphorene. The OPL is determined as $OPL_{BP} = -\frac{\lambda}{2\pi}(\phi_{BP} - \phi_{SiO_2})$, where $\lambda = 535\ nm$ is the wavelength of the light source, and $\phi_{BP}$ and $\phi_{SiO_2}$ are the phase shifts of the light reflected from the phosphorene flake and the SiO$_2$/Si substrate (Figure 1f inset), respectively. The one-to-one relation between the OPL and the layer number is firmly established by a first-principles optical calculation and experimental calibration, as shown in Figure 1f. Note that even though the thickness of one-layer phosphorene is less than 1 nm, its OPL is significantly larger than 20 nm owing to the multiple interfacial light reflections (Supplementary Information). That is, the virtual thickness of a phosphorene flake is amplified by more than 20 times in the optical interferometry, making the flakes easily identified. In the experiment, phase-shifting interferometry (PSI) is used to measure the OPL by analyzing the digitized interference pattern

(Supplementary Information). The step change of the OPL is ~20 nm for each additional phosphorene layer, as indicated by the red dots in Figure 1f. Considering that the accuracy of the instrument is ~0.1 nm, a step change of 20 nm yields an extremely robust measurement. In our experiment, the layer number of the phosphorene flake was first estimated according to its optical contrast in an optical microscope (Figure 1c) and then accurately identified by PSI (Figure 1d and e). After the PSI, the sample was placed into a microscope-compatible chamber for photoluminescence (PL) measurements, with a slow flow of nitrogen gas to prevent the degradation of the sample[7].

Trions, having a many-body bound state, are formed through the interplay between the exciton and carrier. The density of trions can be modulated by controlling the carrier doping level using various methods, such as electrostatic modulation[12,13], and chemical doping[27]. Here, we demonstrate the reversible electrostatic tunability of the exciton charging effects from positive ($A^+$) to neutral ($A°$) in a 3L phosphorene MOS device (Figure 2), using gate-dependent PL measurements. The measured PL spectra exhibit two clear peaks with central wavelengths at ~1100 and ~1320 nm, whose intensities are highly dependent on the back-gate voltage. The higher-energy peak (~1100 nm) is attributed to the exciton emission, and the lower-energy peak (~1320 nm) is due to the trion emission[20]. To show the evolution of the exciton and trion, we applied a gate voltage $V_g$ to pump extra charges into the phosphorene. We used Lorentzian curves to fit the measured PL spectra to extract the exciton and trion spectral components, as indicated by the red and blue curves in Figure 2a, respectively. The Si/SiO$_2$ substrate also has a PL peak at ~1100 nm that is independent of the gate voltages (Figure S1). The emission intensity from the substrate is far weaker than that from phosphorene, and it can be easily separated from the measured voltage-dependent PL spectra (Figure 2a).

At a voltage bias of -50 V, positive charges are pumped into the phosphorene, and almost all excitons become charged trions. As a result, the PL emission from neutral excitons A° at a wavelength of ~1100 nm is absent (Figure 2a). In contrast, the PL emission from positive trions A$^+$ at ~1320 nm is extremely strong. As we gradually depleted the positive charges by changing the voltage from -50 to 50 V, the PL emission from the neutral excitons became increasingly prominent at ~1100 nm, while that of the positive trions became progressively weaker (Figures 2a and b) simultaneously. Such a transfer of the spectral weight is directly caused by the depletion of positive charges, i.e., A$^+$ - h → A°, where h represents a hole. In principle, negative trions can also be expected when $V_g$ is large enough to introduce sufficient electron doping to offset the intrinsic hole doping in the phosphorene layer, which can be realized by replacing the thick oxide layer with thin, high-*k* dielectric materials. The central wavelengths of the emission peaks from A° and A$^+$ are almost independent of the back-gate voltages (Figure 2c), and they exhibit a large energy difference of ~180 meV, which is the trion binding energy. Remarkably, such a large binding energy has not been observed in any other higher-than-one-dimensional material. It is comparable to that from 1D semiconductors[16,17] and is approximately one order of magnitude larger than that from TMD 2D semiconductors[14,15]. Positive trions with a large binding energy have been observed in multiple 3L phosphorene MOS devices (Figure S2).

The ultra-high trion binding energy measured in few-layer phosphorene is caused by the material's unique anisotropic quasi-1D excitonic nature, which can be measured using the linearly polarized emission, as theoretically predicted by Yang et al[9]. Next, we demonstrate the quasi-1D nature of the excitonic effects in few-layer phosphorene using PL measurements with an angle-resolved excitation and emission. Here, we show that the PL emission is completely linearly polarized along the armchair direction of the crystal. Figure 3 presents the results from a representative 2L phosphorene sample, indicating a strong PL emission peak with a central wavelength at 967 nm—the same wavelength value that we reported before[7]. This PL peak was demonstrated to be from the exciton emission in a 2L phosphorene MOS device (Figure S3). In the setup of the angle-resolved PL measurement (Figure 3a), a linearly polarized Nd:YAG laser with a wavelength of 532 nm was used as the excitation source. The polarization angle of the incident light ($\theta_1$) is controlled by an angle-variable half-wave plate. The polarization angle of the emission ($\theta_2$) is characterized by inserting an angle-variable polarizer in front of the detector. $\theta_1$ and $\theta_2$ are relative to the same zero-degree reference, which can be arbitrarily selected in the beginning. In the first characterization of the PL excitation polarization dependence, only the half-wave plate was used, and the polarizer in front of the detector was removed. We found that the PL intensity strongly depends on the excitation polarization angle $\theta_1$ (Figure 3c). This strong PL excitation polarization dependence is due to the highly anisotropic optical absorption in the phosphorene[8,9]. Because of the symmetry in its band structure and the optical selection rules[28], 2L phosphorene strongly absorbs armchair-polarized light and is transparent to zigzag-polarized light with energies between 0.5 and 2.8 eV[9]. According to the polarization-dependent PL excitation, we can determine the crystalline orientation of the phosphorene flake, and then the zigzag direction is selected as the shared zero-degree reference for $\theta_1$ and $\theta_2$ (Figure 3b). Next, we measured the polarization of the PL emission by fixing the excitation polarization angle $\theta_1$ at 90°. We determined the maximum PL intensity at $\theta_2 = 90°$ and the minimum (almost zero) PL intensity at $\theta_2 = 0°$ (Figures 3d and e). Thus, the PL emission is completely linearly polarized, and the emission polarization is along the armchair direction. The measured linear dichroism (*LD*) is close to unity. *LD* is defined as $LD = (I_x - I_y)/I_x$, where $I_x$ and $I_y$ are the PL emission peak intensities along the armchair and zigzag directions, respectively. The central wavelength of the PL emission peak is independent of the emission polarization angle (Figure 3e). Then, we set the excitation angle $\theta_1$ to an arbitrarily value (e.g., 15°) and measured the polarization of PL emission again. We found that the PL emission was still completely polarized along the armchair direction (Figure 3f). The polarization of the PL emission is independent of the excitation and is determined by the intrinsic properties of phosphorene[28].

Few-layer phosphorene has quasi-1D excitons, as we demonstrated, while the carrier in phosphorene is still in a 2D space with a slightly anisotropic carrier mobility[5]. Similarly to the exciton emission, the emission of the charged excitons is linearly polarized along the armchair direction, as demonstrated by the linearly polarized PL emission from the trions ($A^+$) in 3L phosphorene (Figure 4a). The measured *LD* value of this trion emission is ~0.9, which is less than that for the polarized exciton emission. This reduced *LD* value of the trion emission is attributed to the influence of the 2D carrier in few-layer phosphorene.

Lastly, the exciton binding energy is predicted to decrease as the stacking layer number increases, owing to the van der Waals interactions between the neighboring sheets[9,29]. The trion binding energy is estimated as a fraction of the exciton binding energy in a 2D system[29], which leads to the layer-sensitive trion behavior. In order to confirm the aforementioned predictions, we measured the trion binding energies in 2L and 4L phosphorene samples (Figures S3, S4 and S5) and determined that the measured trion binding energy decreases as the layer number increases (two to four layers) (Figure 4b). For each layer number, we characterized two devices; the average trion binding energies in 2L, 3L, and 4L phosphorene are 190, 160, and 90 meV, respectively. We also used the theoretical model of trions developed by Thilagam[29], combining the empirical equation of the exciton binding energy by Yang et al.[9] and the carrier effective masses calculated by Ji et al.[8] to calculate the trion binding energies in few-layer phosphorene (Supplementary Information). The theoretical calculation results agree well with our experimental observations.

In conclusion, we observed extraordinarily bound quasi-1D trions in 2D phosphorene atomic semiconductor crystals. The measured ultra-high trion binding energies in few-layer phosphorene are approximately one order of magnitude higher than those in 2D TMD semiconductors. The large trion binding energy is due to the strongly confined 1D excitonic nature in few-layer phosphorene, which is demonstrated by our measured linearly polarized PL emission. Phosphorene possesses both a large optical cross-section, as typically exhibited by a 2D material system, and high trion binding energies, as typically exhibited by a 1D system, allowing remarkable optoelectronic applications, including tunable light sources, photo-detectors and spin manipulation devices. In addition, the measured layer-dependent trion binding energies in few-layer phosphorene are consistent with our theoretical calculations, which will enable a wide range of tunable energy gaps and corresponding optoelectronic applications. Few-layer phosphorene also serves as a room-temperature platform for investigating many-body interactions and excitonic physics.

**Methods**

**Device Fabrication and Characterization.** We used mechanical exfoliation to drily transfer[26] a phosphorene flake onto a $SiO_2$/Si substrate (275 nm thermal oxide on $n^+$-doped silicon), near a pre-patterned gold electrode. The gold electrodes were patterned by conventional photolithography, metal deposition, and lift-off processes. Another thick graphite flake was similarly transferred to electrically bridge the phosphorene flake and the gold electrode, forming a MOS device. All PL and polarization measurements were conducted using a T64000 micro-Raman system equipped with a charge-coupled device (CCD) and InGaAs detectors, along with a 532nm Nd:YAG laser as the excitation source. For all measurements, the sample was placed into a microscope-compatible chamber with a slow flow of protection nitrogen gas to prevent the degradation of the sample. To avoid laser-induced sample damage, all PL spectra were recorded at low power levels: P ~20 μW. For the PL measurements, an integration time of 30 s was used. The electrical bias was applied using a Keithley 4200 semiconductor analyzer.

**Numerical Simulation.** Stanford Stratified Structure Solver (S4)[30] was used to calculate the phase delay. The method numerically solves Maxwell's equations in multiple layers of structured materials by expanding the field in the Fourier-space.

**Author Contributions**

Y. R. L. designed the project; S. Z. did the PL and Raman measurements and data analysis; R. J. X. conducted the device fabrication, PL data fitting analysis, and part of the theoretical calculations; S. Z. and R. J. X. conducted the PSI measurements; J. Y., J. J. P., and Y. W. M. contributed to sample preparation; B. B. X. performed the image post-processing; F. W. and L. F. built the optical characterization setup; Z. W. and Z. F. Y performed the theoretical simulations for the OPL of phosphorene. All authors contributed to the manuscript.

**Acknowledgements**

We wish to acknowledge support from the ACT node of the Australian National Fabrication Facility (ANFF). We also thank Professor Chennupati Jagadish, Professor Hoe Tan, and Professor Barry Luther-Davies from the Australian National University for facility support, as well as Dr. Gang Zhang from the Institute of High Performance Computing, A*STAR, Singapore, for the helpful discussion. We acknowledge financial support from an ANU PhD scholarship, the Office of Naval Research (USA) (grant number N00014-14-1-0300), the Australian Research Council (grant number DE140100805), and the ANU Major Equipment Committee.

**Competing Financial Interests**

The authors declare that they have no competing financial interests.

**References**

1. Liu, H. *et al.* Phosphorene: An unexplored 2D semiconductor with a high hole mobility. *ACS Nano* **8**, 4033-4041 (2014).
2. Buscema, M. *et al.* Fast and broadband photoresponse of few-layer black phosphorus field-effect transistors. *Nano Lett.* **14**, 3347-3352 (2014).
3. Hong, T. *et al.* Polarized photocurrent response in black phosphorus field-effect transistors. *Nanoscale* (2014).
4. Fei, R. & Yang, L. Strain-engineering the anisotropic electrical conductance of few-layer black phosphorus. *Nano Lett.* **14**, 2884-2889 (2014).
5. Xia, F., Wang, H. & Jia, Y. Rediscovering black phosphorus as an anisotropic layered material for optoelectronics and electronics. *Nat. Commun.* **5** (2014).
6. Li, L. *et al.* Black phosphorus field-effect transistors. *Nat. Nanotechnol.* **9**, 372-377 (2014).
7. Zhang, S. *et al.* Extraordinary photoluminescence and strong temperature/angle-dependent raman responses in few-layer phosphorene. *ACS Nano* **8**, 9590-9596 (2014).
8. Qiao, J., Kong, X., Hu, Z.-X., Yang, F. & Ji, W. High-mobility transport anisotropy and linear dichroism in few-layer black phosphorus. *Nat. Commun.* **5** (2014).
9. Tran, V., Soklaski, R., Liang, Y. & Yang, L. Layer-controlled band gap and anisotropic excitons in few-layer black phosphorus. *Phys. Rev. B* **89**, 235319 (2014).
10. Geim, A. K. & Novoselov, K. S. The rise of graphene. *Nat. Mater.* **6**, 183-191 (2007).
11. Radisavljevic, B., Radenovic, A., Brivio, J., Giacometti, V. & Kis, A. Single-layer $MoS_2$ transistors. *Nat. Nanotechnol.* **6**, 147-150 (2011).
12. Mak, K. F. *et al.* Tightly bound trions in monolayer $MoS_2$. *Nat. Mater.* **12**, 207-211 (2013).
13. Ross, J. S. *et al.* Electrical control of neutral and charged excitons in a monolayer semiconductor. *Nat. Commun.* **4**, 1474 (2013).
14. Huard, V., Cox, R. T., Saminadayar, K., Arnoult, A. & Tatarenko, S. Bound states in optical absorption of semiconductor quantum wells containing a two-dimensional electron gas. *Phys. Rev. Lett.* **84**, 187-190 (2000).
15. Bracker, A. S. *et al.* Binding energies of positive and negative trions: From quantum wells to quantum dots. *Phys. Rev. B* **72**, 035332 (2005).
16. Matsunaga, R., Matsuda, K. & Kanemitsu, Y. Observation of charged excitons in hole-doped carbon nanotubes using photoluminescence and absorption spectroscopy. *Phys. Rev. Lett.* **106**, 037404 (2011).
17. Park, J. S. *et al.* Observation of negative and positive trions in the electrochemically carrier-doped single-walled carbon nanotubes. *J. Am. Chem. Soc.* **134**, 14461-14466 (2012).
18. Klimov, V. I. *et al.* Optical gain and stimulated emission in nanocrystal quantum dots. *Science* **290**, 314-317 (2000).
19. Scholes, G. D. & Rumbles, G. Excitons in nanoscale systems. *Nat. Mater.* **5**, 683-696 (2006).
20. Kheng, K. *et al.* Observation of negatively charged excitons in semiconductor quantum wells. *Phys. Rev. Lett.* **71**, 1752-1755 (1993).
21. Wigner, E. On the interaction of electrons in metals. *Phys. Rev.* **46**, 1002-1011 (1934).
22. Galland, C. & Imamoğlu, A. All-optical manipulation of electron spins in carbon-nanotube quantum dots. *Phys. Rev. Lett.* **101**, 157404 (2008).
23. Xu, X., Yao, W., Xiao, D. & Heinz, T. F. Spin and pseudospins in layered transition metal dichalcogenides. *Nat. Phys.* **10**, 343-350 (2014).

24  Carter, S. G. *et al.* Quantum coherence in an optical modulator. *Science* **310**, 651-653 (2005).
25  High, A. A., Novitskaya, E. E., Butov, L. V., Hanson, M. & Gossard, A. C. Control of exciton fluxes in an excitonic integrated circuit. *Science* **321**, 229-231 (2008).
26  Castellanos-Gomez, A. *et al.* Deterministic transfer of two-dimensional materials by all-dry viscoelastic stamping. *2D Materials* **1**, 011002 (2014).
27  Mouri, S., Miyauchi, Y. & Matsuda, K. Tunable photoluminescence of monolayer $MoS_2$ via chemical doping. *Nano Lett.* **13**, 5944-5948 (2013).
28  Yuan, H. *et al.* Broadband linear-dichroic photodetector in a black phosphorus vertical p-n junction. *arXiv:1409.4729* (2014).
29  Thilagam, A. Two-dimensional charged-exciton complexes. *Phys. Rev. B* **55**, 7804-7808 (1997).
30  Liu, V. & Fan, S. S4 : A free electromagnetic solver for layered periodic structures. *Comput. Phys. Commun.* **183**, 2233-2244 (2012).

**FIGURE CAPTIONS**

**Figure 1 | Phosphorene characteristics and devices. a,** Schematic plot of phosphorene layer structure. **b,** Schematic plot of a phosphorene MOS device. **c,** Optical microscope image of the MOS device with bi-layer phosphorene (labeled as "2L"). **d,** Phase shifting interferometry (PSI) image of the region inside the box indicated by the dashed line in **(c)**. **e,** PSI measured OPL values versus position for 2L phosphorene along the dashed line in **(d)**. **f,** Statistical data of the OPL experimental values from PSI for 1–6L phosphorene samples. For each layer number of phosphorene, at least three different samples were characterized for the statistical measurements. Theoretical simulation data is also plotted, for comparison. Inset: schematic plot indicating the PSI measured phase shifts of the reflected light from the phosphorene flake ($\phi_{BP}$) and the SiO$_2$ substrate ($\phi_{SiO_2}$).

**Figure 2 | Gate dependence of the exciton and trion in a 3L phosphorene MOS device. a,** Measured photoluminescence (PL) spectra (solid dark grey lines) under various back-gate voltages. PL spectra are fit to Lorentzians (solid red lines are the exciton components, solid blue lines are the trion components, solid light grey lines are the Si components, and dashed pink lines are the cumulative results for the fitting). **b,** PL intensity of exciton and trion as a function of gate voltage. Inset: representation of the dissociation of a trion into an exciton and a hole at the Fermi level. **c,** Peak energy of exciton and trion as a function of gate voltage.

**Figure 3 | Quasi-1D exciton in phosphorene. a,** Schematic plot of setup for measurement to characterize the polarization dependence of PL excitation and emission. The polarization angle ($\theta_1$) incident excitation light is controlled by an angle-variable half-wave plate, and the polarization angle ($\theta_2$) of the PL emission is characterized by inserting an angle-variable polarizer in front of the detector. **b,** Schematic plot showing top view of phosphorene lattice structure and coordinates for polarization angles $\theta_1$ and $\theta_2$. **c,** Measured excitation polarization

dependence of PL peak intensities from 2L phosphorene. For this measurement, the polarizer in front of the detector was removed. **d,** Measured emission polarization dependence of PL peak intensities from the same 2L phosphorene, with excitation angle of 90°. **e,** Measured emission polarization dependence of PL spectra, with excitation angle of 90°. **f,** Measured emission polarization dependence of PL peak intensities from the same 2L phosphorene, with excitation angle of 15°. Red lines in c, d, and f are the fitting curves.

**Figure 4 | Quasi-1D trion and layer-dependent trion binding energies in few-layer phosphorene. a,** Measured polarization dependence of the trion ($A^+$) emission measured from 3L phosphorene, with excitation angle of 95°. **b,** Measured layer-dependent trion binding energies in few-layer phosphorene (two to four layers). Theoretical trion binding energy is presented for comparison.

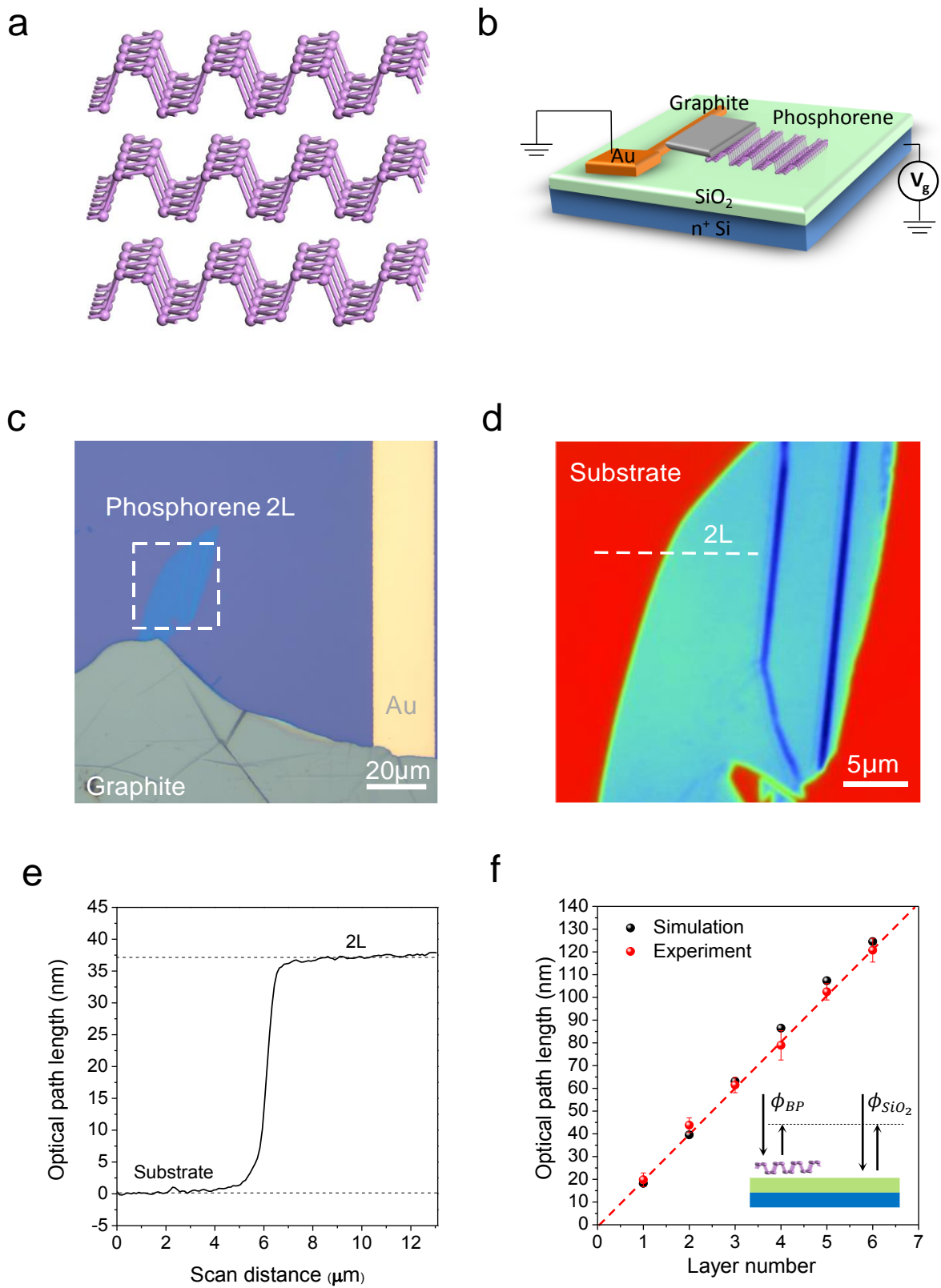

Figure 1

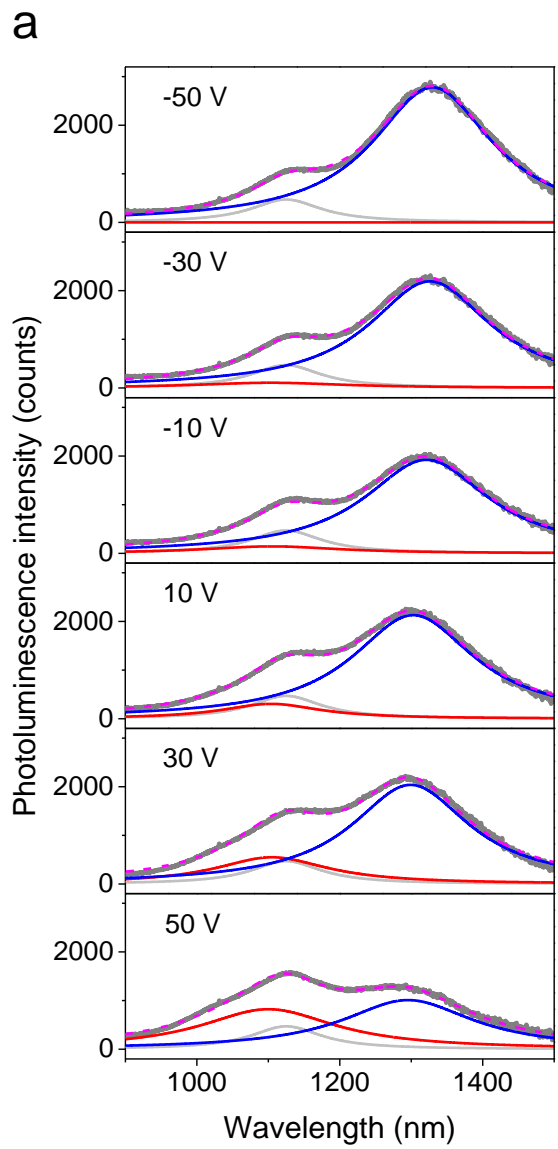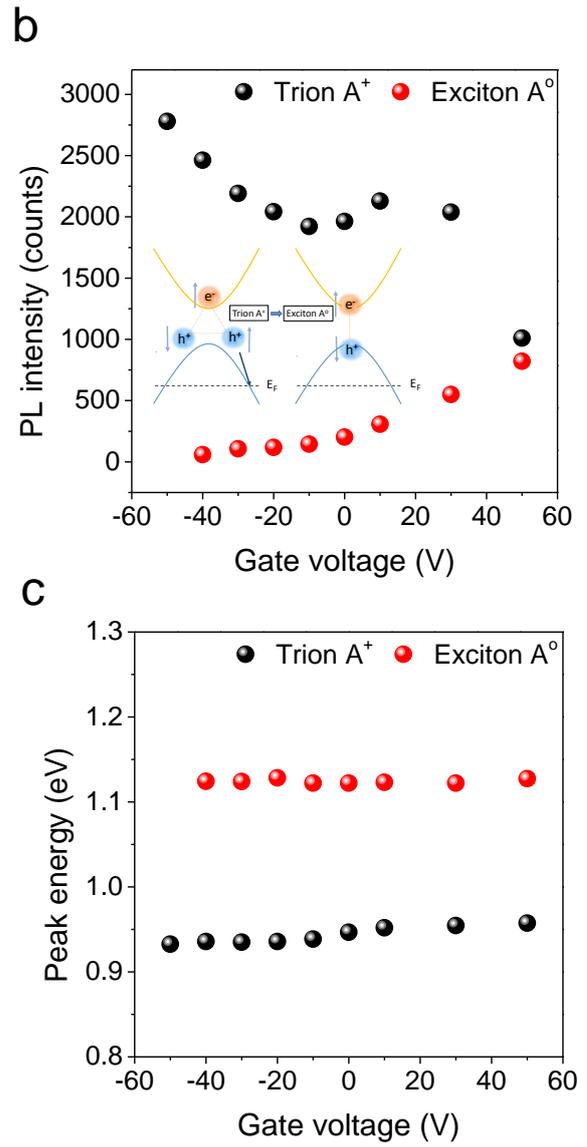

Figure 2

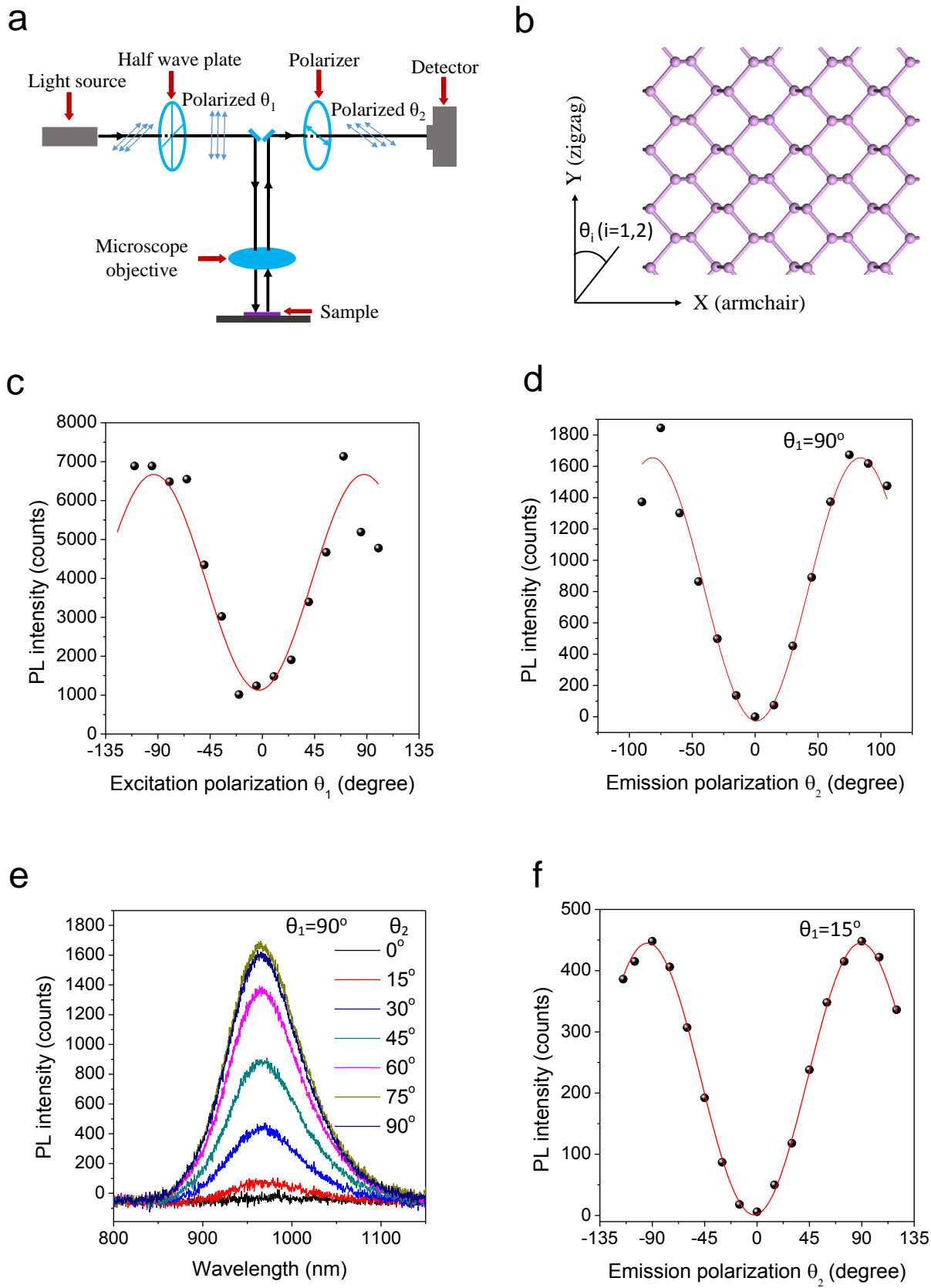

Figure 3

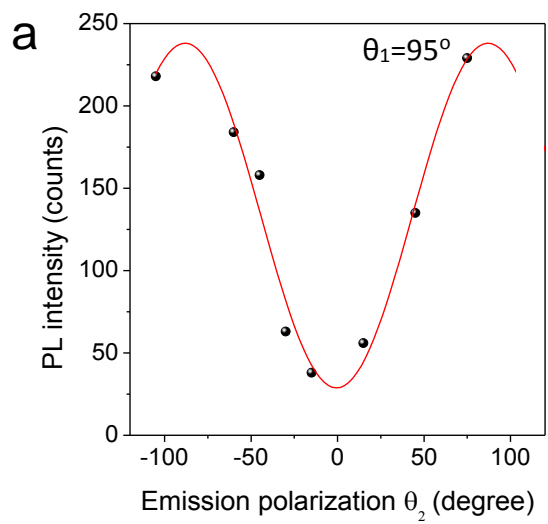 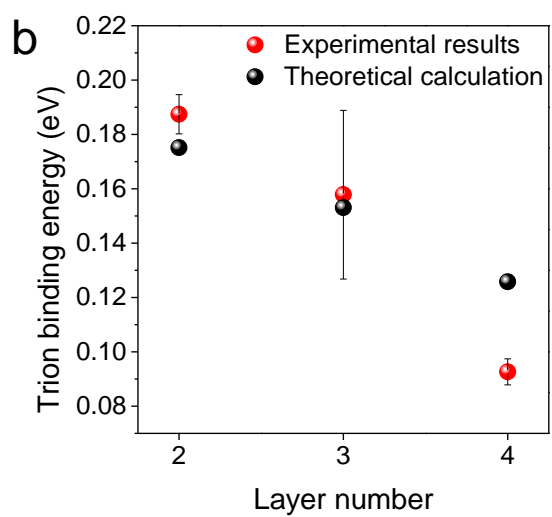

Figure 4

# Supplementary Information for

# Extraordinarily bound quasi-one-dimensional trions in two-dimensional phosphorene atomic semiconductors

Shuang Zhang,[1†] Renjing Xu,[1†] Fan Wang,[2] Jiong Yang,[1] Zhu Wang,[3] Jiajie Pei,[1] Ye Win Myint,[1]

Bobin Xing,[1] Zongfu Yu,[3] Lan Fu,[2] Qinghua Qin,[1] and Yuerui Lu[1*]

[1]Research School of Engineering, College of Engineering and Computer Science, the Australian National University, Canberra, ACT, 0200, Australia

[2]Department of Electronic Materials Engineering, Research School of Physics and Engineering, the Australian National University, Canberra, ACT, 0200, Australia

[3]Department of Electrical and Computer Engineering, University of Wisconsin, Madison, Wisconsin 53706, USA

[†] These authors contributed equally to this work

**\*** To whom correspondence should be addressed: Yuerui Lu (yuerui.lu@anu.edu.au)

1. **Photoluminescence (PL) measurements from various few-layer (two to four layers) phosphorene devices**

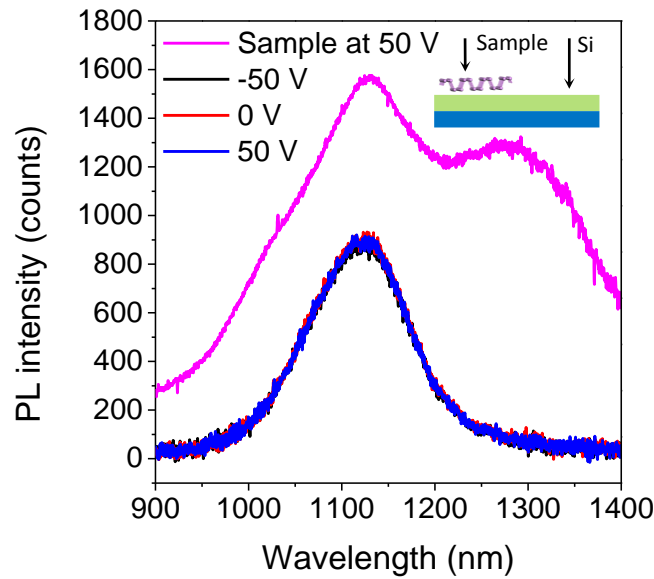

**Figure S1 | Photoluminescence (PL) from the SiO$_2$/Si substrate used for the MOS device under different back-gate voltages**. PL spectrum of the 3L phosphorene MOS at gate voltage of 50 V (shown in Figure 2a) is also plotted here for comparison. Insert: Schematic of the measurement positions. The PL measurement conditions are kept the same for the phosphorene sample and the SiO$_2$/Si substrate.

In Figure S1, the PL spectra measured from the SiO$_2$/Si substrate at back-gate voltages of -50, 0 and 50 V are almost identical, originating from the voltage independent Si PL at ~ 1100 nm; whereas the PL peak intensity at ~1100 nm from the 3L phosphorene MOS device under back-gate voltage 50 V is even higher than that from the substrate. Based on this, we believe that the ~1100 nm PL peak on 3L phosphorene comes from both the phosphorene and Si substrate. The emission intensity from the substrate is much weaker than that from phosphorene and it can be easily separated from the measured voltage-dependent PL spectra (Figure 2a).

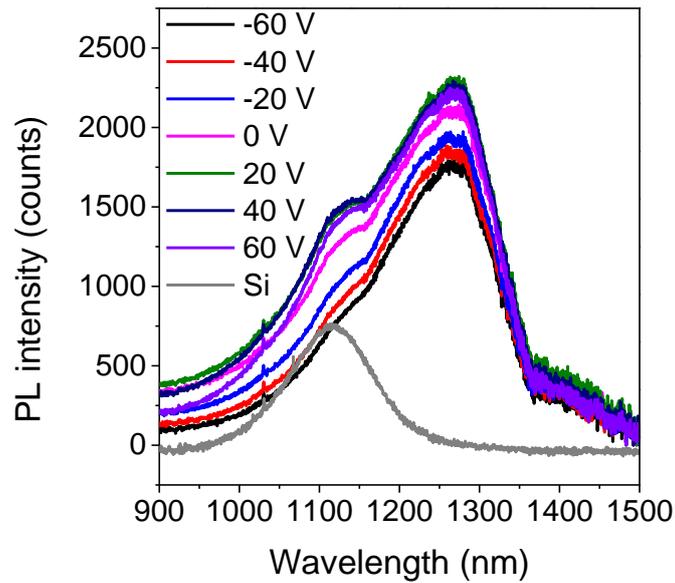

**Figure S2 | Gate dependence of the exciton and trion in another 3L phosphorene MOS device.** The Si PL spectrum measured from the substrate of this device is also shown for comparison. Note that the Si PL peak intensity here is slightly different from the one shown in Figure S1, due to a slight difference of the laser excitation power used in these two measurements.

Figure S2 shows the repeatable gate dependence of the exciton and trion in another 3L phosphorene MOS device. The measured PL spectra show two clear peaks with central wavelengths at ~1100 nm and ~1260 nm, respectively, whose intensities strongly depend on the applied back-gate voltage $V_g$. When $V_g$ changes from -60 to 60 V, the peak intensity of the ~1100 nm PL peak increases, which is consistent with the trend shown in Figure 2a. Here, the difference between peak energies of $A^o$ (~1100 nm) and $A^-$ (~1260 nm) is ~130 meV. We believe that the trion binding energy also hightly depends on the quality of the phosphorene in the MOS device, i.e. the initial doping level and deficiencies.

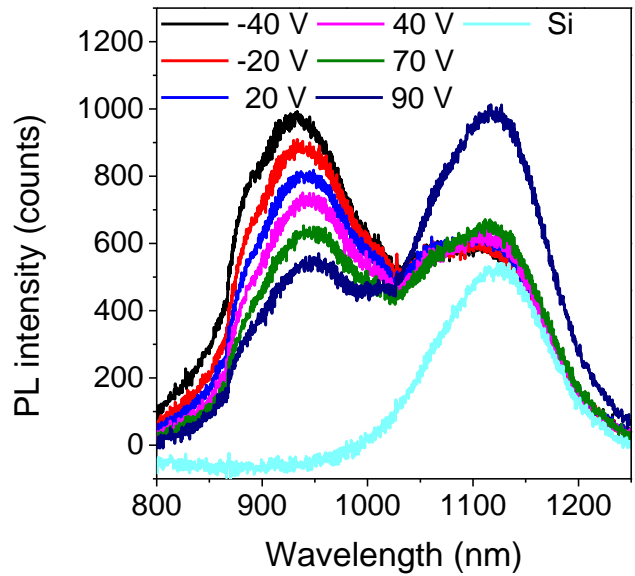

**Figure S3 | Gate dependence of the exciton and trion in a 2L phosphorene MOS device.** Measured PL spectra in the range of 800 to 1,250 nm under different back-gate voltages. The Si PL spectrum obtained from the $SiO_2$/Si substrate is shown for comparison.

Through gate dependent PL measurements, we also demonstrate the reversible electrostatic tunability of the exciton charging effects from negative ($A^-$) to neutral ($A^o$) in a 2L phosphorene MOS device (Figure S3). The measured PL spectra show two clear peaks with central wavelengths at ~930 nm and ~1120 nm respectively, whose intensities strongly depend on the applied back-gate voltage $V_g$. When $V_g$ changes from -40 to 90 V, the intensity of the ~930 nm PL peak decreases, while that of the ~1120 nm PL peak increases (Figure S3). $SiO_2$/Si substrate also shows a PL peak at ~1100 nm, but the measured PL intensity on pure Si substrate is independent of $V_g$ and its intensity is even smaller than the intensity of the ~1120 nm PL peak on 2L phosphorene. Based on this, not to mention the absorption of Si PL from the phosphorene layer, we believe that the ~1120 nm PL peak on 2L phosphorene comes from both the phosphorene and $SiO_2$/Si substrate and the $V_g$ induced intensity enhancement at ~1120 nm comes from the phosphorene only. The higher-energy phosphorene emission at ~930 nm is the neutral exciton, A⁰, and the lower-energy peak at ~1120 nm is a trion. The exction spectral weight is transferred to the trion as negative charges are injected to the phosphorene ($V_g$ changes from -40 to 90V), which indicates the trion here is negatively charged ($A^-$). The conversion from $A^o$ to $A^-$ can be represented as $A^o + e \rightarrow A^-$, where e represents an electron. The difference between peak energy of $A^o$ (~930 nm) and $A^-$ (~1120 nm) is ~190 meV.

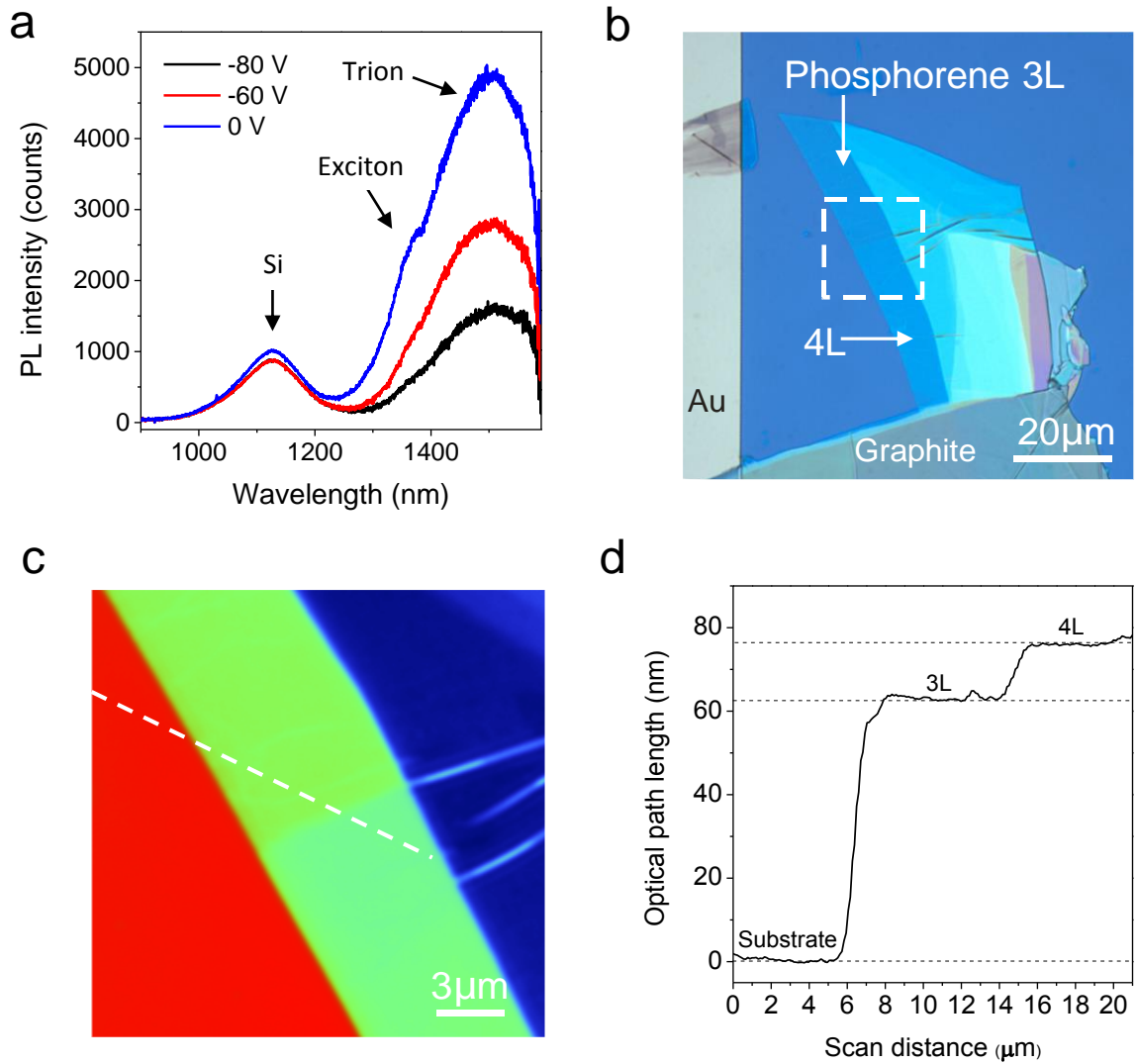

**Figure S4 | Exciton and trion in a 4L phosphorene MOS device. a,** Measured PL spectra from a 4L phoshorene MOS device, with back-gate voltage bias. The corresponding peaks for Si, exciton and trion are indicated. **b,** Optical microscope image of the phosphorene MOS device. **c,** PSI image of the phosphorene flake from the dash line box area indicated in (b). **d,** Measured OPL values versus position along the dash line in (c).

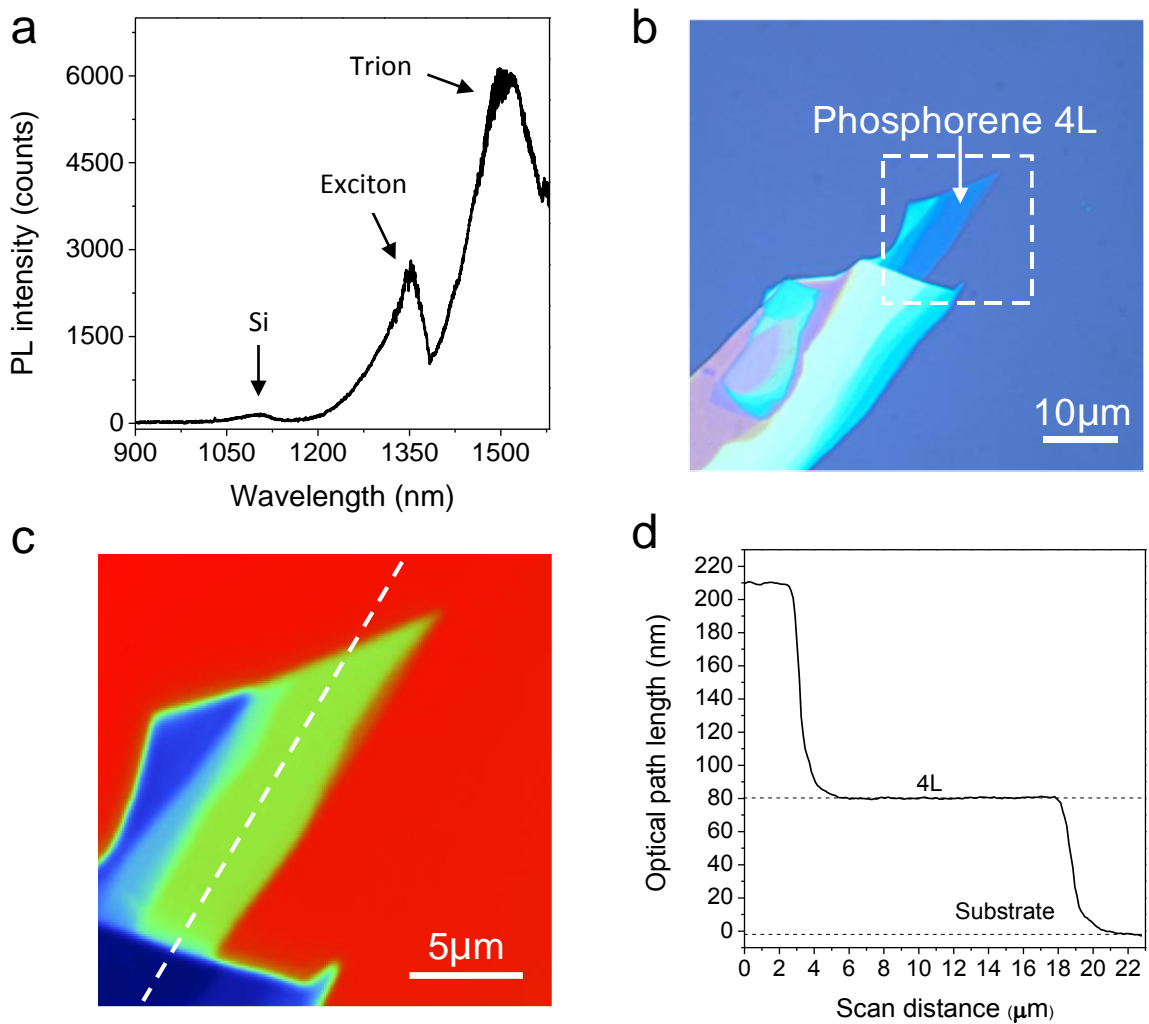

**Figure S5 | Exciton and trion in another 4L phosphorene. a,** Measured PL spectra from a 4L phoshorene, without back gate voltage bias. The corresponding peaks for Si, exciton and trion are indicated. **b,** Optical microscopy image of 4L phosphorene accompanied with data in (a). **c,** PSI image of the phosphorene flake from the dash line box area indicated in (b). **d**, Measured OPL values versus position along the dash line in (c).

In both Figure S4 and S5, the measured PL spectra show two clear peaks with central wavelengths at ~1350 nm and ~1500 nm respectively. In Figure S4, the PL intensities strongly depend on the applied back-gate voltage $V_g$. The higher-energy phosphorene emission at ~1350 nm is the neutral exciton, A⁰, and the lower-energy peak at ~1500 nm is from trion emission. The difference between peak energy of A⁰ (~1350 nm) and A⁻ (~1500 nm) is ~90 meV.

Without any back-gate bias, the PL spectrum in Figure S5 shows two peaks, of which the high-energy one is from exciton and the low-energy one is from trion. The central wavelengths of these two peaks are the same with those in Figure S4. We found that the spectral weight of the emission peaks from trion and excition are highly related to the initial doping level of the phosphorene samples.

## 2. Trion binding energy calculation

Our angle-resolved PL measurements on few-layer phosphorene samples clearly indicate that the PL emission from trion and excitons are both linearly polarized along the armchair direction. Hence we calculate the trion binding energy of few-layer phosphorene in the armchair direction. The binding energy of both negatively and positively charged trions in 2D system can be estimated using following equations[1]:

$$Eb_{A^-} = \left[\left(\frac{3}{2}\right)^2 \left(1 + \frac{2\sigma+1}{\sigma^2+4\sigma+2}\right)^{-1} - 1\right] \frac{1}{1+\sigma} * \frac{m_e^* Eb_{ex}}{\gamma M_R} \quad (S1)$$

$$Eb_{A^+}(\sigma) = \frac{Eb_{A^-}\left(\frac{1}{\sigma}\right)}{\sigma} \quad (S2)$$

where $Eb_{A^-}$ and $Eb_{A^+}$ are the binding energy of negatively and positively charged trions, respectively, γ is a measure of dimensionality, $M_R$ is the reduced mass of $m_e^*$ and $m_h^*$, $Eb_{ex}$ is the exciton binding energy and $\sigma = m_e^*/m_h^*$ is the ratio of effective mass between electrons and holes.

The exciton binding energy in few-layer phosphorene has been predicted by Yang *et al*[2] as:

$$Eb_{ex} = \frac{A}{N^\alpha} + B \tag{S3}$$

where $\alpha = 0.53$, $A = 0.83$ eV, $B = 0.03$ eV and $N$ is phosphorene layer number.

**Table 1 | Carrier effective masses in armchair direction in few-layer phosphorene[3]. $m_X^*$ and $m_0$ represent effective mass in armchair direction and rest mass of carriers, respectively.**

| Carrier type | Layer number | $m_X^*/m_0$ |
|---|---|---|
| Electron | 1 | 0.17 |
|  | 2 | 0.18 |
|  | 3 | 0.16 |
|  | 4 | 0.16 |
|  | 5 | 0.15 |
| Hole | 1 | 0.15 |
|  | 2 | 0.15 |
|  | 3 | 0.15 |
|  | 4 | 0.14 |
|  | 5 | 0.14 |

**Table 2 | Trion binding energy in few-layer phosphorene. $Eb_{ex}$ represents the binding energy of neutral excitons.**

| Trion type | Layer number | $Eb_{ex}$ (eV) | $Eb_{A^+}$ or $Eb_{A^-}$ (eV) |
|---|---|---|---|
| Negatively charged trions ($A^-$) | 1 | 0.860 | 0.252 |
|  | 2 | 0.605 | 0.179 |
|  | 3 | 0.494 | 0.143 |
|  | 4 | 0.428 | 0.126 |
|  | 5 | 0.384 | 0.111 |
| Positively Charged trions ($A^+$) | 1 | 0.860 | 0.286 |
|  | 2 | 0.605 | 0.215 |
|  | 3 | 0.494 | 0.153 |
|  | 4 | 0.428 | 0.144 |
|  | 5 | 0.384 | 0.120 |

Ji *et al*[3] had calculated the effective masses of both electrons and holes in phosphorene along armchair direction and the results are presented in Table 1. Using equations S1-S3 and the effective mass values from Table 1, the binding energies of neutral excitons and charged trions in few-layer phosphorene are calculated and presented in Table 2.

## 3. Phase-shifting interferometry (PSI) working principle

PSI was used to investigate the surface topography based on analyzing the digitized interference data obtained during a well-controlled phase shift introduced by the Mirau interferometer[4]. The PSI system (Vecco NT9100) used in our experiments operates with a green LED source centered near 535 nm by a 10 nm band-pass filter[5]. The schematic of the PSI system is shown in Figure S6.

The working principle of the PSI system is as follows[6]. For simplicity, wave front phase will be used for analysis. The expressions for the reference and test wave-fronts in the phase shifting interferometer are:

$$w_r(x,y) = a_r(x,y)e^{i\phi_r(x,y)} \quad (S4)$$

$$w_t(x,y,t) = a_t(x,y)e^{i[\phi_t(x,y)+\delta(t)]} \quad (S5)$$

where $a_r(x,y)$ and $a_t(x,y)$ are the wavefront amplitudes, $\phi_r(x,y)$ and $\phi_t(x,y)$ are the corresponding wavefront phases, and $\delta(t)$ is a time-dependent phase shift introduced by the Mirau interferometer. $\delta(t)$ is the relative phase shift between the reference beam and the test beam.

The interference pattern of these two beams is:

$$w_i(x,y,t) = a_r(x,y)e^{i\phi_r(x,y)} + a_t(x,y)e^{i[\phi_t(x,y)+\delta(t)]} \quad (S6)$$

The interference intensity pattern detected by the detector is:

$$I_i(x,y,t) = w_i^*(x,y,t) * w_i(x,y,t) = I'(x,y) + I''(x,y)\cos[\phi(x,y) + \delta(t)] \quad (S7)$$

where $I'(x,y) = a_r^2(x,y) + a_t^2(x,y)$ is the averaged intensity, $I''(x,y) = 2a_r(x,y) * a_t(x,y)$ is known as intensity modulation and $\phi(x,y)$ is the wavefront phase shift $\phi_r(x,y) - \phi_t(x,y)$.

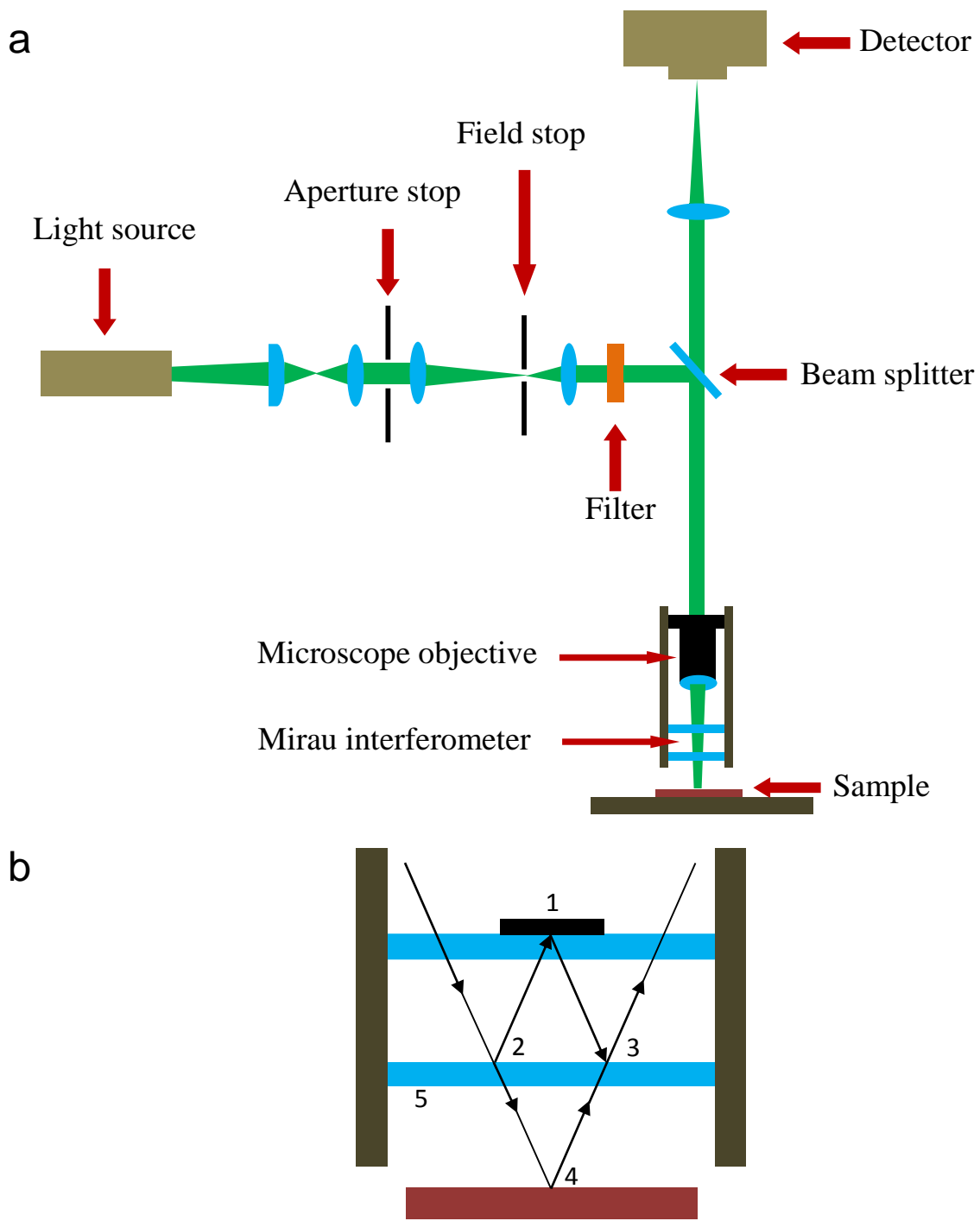

**Figure S6 | Schematic plot of the phase shifting interferometry (PSI) system. a,** Schematic plot of the PSI system. **b,** Zoomed view of the Mirau interferometer. 1. Reference mirror; 2. First reflection of the reference beam; 3. Third reflection of the reference beam; 4. Reflection of the test/objective beam; 5. Semi-transparent mirror. 2-1-3 represents the reference beam and 2-4-3 represents the test/objective beam.

From the above equation, a sinusoidally-varying intensity of the interferogram at a given measurement point as a function of $\delta(t)$ is shown below:

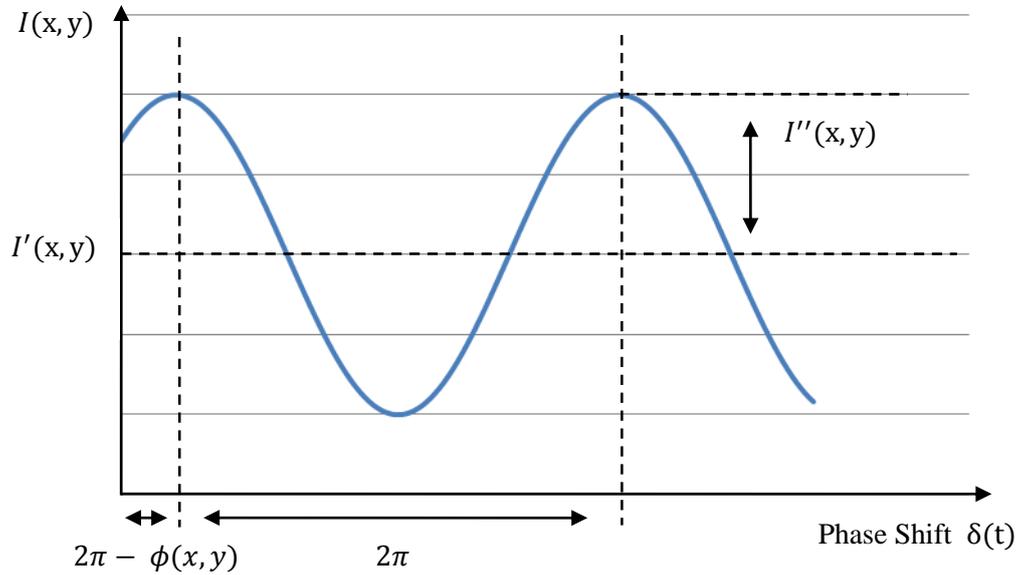

**Figure S7 | Variation of intensity with the reference phase at a point in an interferogram.** $I'(x, y)$ is the averaged intensity, $I''(x, y)$ is half of the peak-to-valley intensity modulation and $\phi(x, y)$ is the temporal phase shift of this sinusoidal variation.

$\delta(t)$ is introduced by the Mirau interferometer, which is shown in Figure S6. When the Mirau interferometer gradually moves toward the sample platform, the optical path length (OPL) of the test beam decreases while the OPL of the reference beam remains invariant.

The computational method of PSI is a four-step algorithm, which needs to acquire four separately recorded and digitalized interferograms of the measurement region. For each separate and sequential recorded interferograms, the phase shift difference is:

$$\delta(t_i) = 0, \frac{\pi}{2}, \pi, \frac{3\pi}{2}; \quad i = 1,2,3,4 \tag{S8}$$

Substituting these four values into the equation S7, leads to the following four equations describing the four measured intensity patterns of the interferogram:

$$I_1(x,y) = I'(x,y) + I''(x,y)\cos[\phi(x,y)] \tag{S9}$$

$$I_2(x,y) = I'(x,y) + I''(x,y)\cos[\phi(x,y) + \tfrac{\pi}{2}] \tag{S10}$$

$$I_3(x,y) = I'(x,y) + I''(x,y)\cos[\phi(x,y) + \pi] \tag{S11}$$

$$I_4(x,y) = I'(x,y) + I''(x,y)\cos[\phi(x,y) + \tfrac{3\pi}{2}] \tag{S12}$$

After the trigonometric identity, this yields:

$$I_1(x,y) = I'(x,y) + I''(x,y)\cos[\phi(x,y)] \tag{S13}$$

$$I_2(x,y) = I'(x,y) - I''(x,y)\sin[\phi(x,y)] \tag{S14}$$

$$I_3(x,y) = I'(x,y) - I''(x,y)\cos[\phi(x,y)] \tag{S15}$$

$$I_4(x,y) = I'(x,y) + I''(x,y)\sin[\phi(x,y)] \tag{S16}$$

The unknown variables $I'(x,y)$, $I''(x,y)$ and $\phi(x,y)$ can be solved by only using three of the four equations; but for computational convenience, four equations are used here. Subtracting equation S14 from equation S16, we have:

$$I_4(x,y) - I_2(x,y) = 2I''(x,y)\sin[\phi(x,y)] \tag{S17}$$

And subtract equation S15 from equation S13, we get:

$$I_1(x,y) - I_3(x,y) = 2I''(x,y)\cos[\phi(x,y)] \tag{S18}$$

Taking the ratio of equation S17 and equation S18, the intensity modulation $I''(x,y)$ will be eliminated as following:

$$\frac{I_4(x,y) - I_2(x,y)}{I_1(x,y) - I_3(x,y)} = \tan[\phi(x,y)] \tag{S19}$$

Rearranging equation S19 to get the wave-front phase shift term $\phi(x,y)$:

$$\phi(x,y) = \tan^{-1}\frac{I_4(x,y)-I_2(x,y)}{I_1(x,y)-I_3(x,y)} \tag{S20}$$

This equation is performed at each measurement point to acquire a map of the measured wavefront. Also, in PSI, the phase shift is transferred to the surface height or the optical path difference (OPD):

$$h(x,y) = \frac{\lambda\phi(x,y)}{4\pi} \tag{S21}$$

$$OPD(x,y) = \frac{\lambda\phi(x,y)}{2\pi} \tag{S22}$$

Here, the OPL of the phosphorene flake $OPL_{BP}$ is calculated as:

$$OPL_{BP} = -(OPD_{BP} - OPD_{SiO_2}) = -\frac{\lambda}{2\pi}(\phi_{BP} - \phi_{SiO_2}) \tag{S23}$$

where $\lambda$ is the wavelength of the light source, $\phi_{BP}$ and $\phi_{SiO_2}$ are the measured phase shifts of the reflected light from the phosphorene flake and the $SiO_2$ substrate, respectively. In our experiments, $\phi_{SiO_2}$ was typically set to be zero, as shown in Figure 1f.

# 4. Calculations for the optical path length (OPL) of atomically thin 2D materials

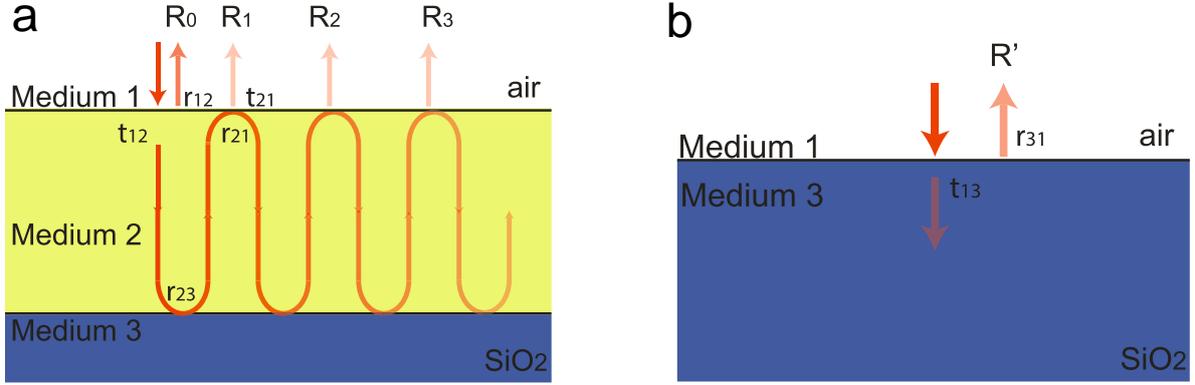

**Figure S8 | a**, Reflection of a three-layer structure. Medium 1 is air, Medium 2 is the 2D material and Medium 3 is an infinite SiO$_2$ substrate. **b**, The reference configuration. Light is incident from air into infinite SiO$_2$ substrate.

The incident light comes from the air resonates inside the 2D material. The total reflection is determined by the interference of all reflected beams $R_i$. To calculate the amplitude of the total reflection, we use $r_{ij}$ ($i,j$=1,2,3) to represent the reflection coefficients when light goes from medium $i$ to medium $j$.

$$r_{ij} = \frac{n_i - n_j}{n_i + n_j} \tag{S24}$$

We use $t_{ij}$ ($i,j$=1,2,3) to represent the transmission from medium $i$ to medium $j$

$$t_{ij} = \frac{2n_i}{n_i + n_j} \tag{S25}$$

where $n_i$, $n_j$ ($i,j$=1,2,3) is the refractive index of medium $i,j$. Assuming that the thickness of the 2D material is $d$ and wave vector of incident light in air is $k_0$, we can calculate the reflection of each order,

$$R_0 = r_{12}$$

$$R_1 = t_{12}r_{23}t_{21}e^{i2k_0nd}$$

$$R_2 = t_{12}r_{23}r_{21}r_{23}t_{21}(e^{i2k_0nd})^2$$

$$R_3 = t_{12}r_{23}r_{21}r_{23}r_{21}r_{23}t_{21}(e^{i2k_0nd})^3 \tag{S26}$$

where $2k_0nd$ is the round trip propagation phase and $n$ is the refractive index of the 2D material.

Then the total reflected amplitude is the summation of all reflections, which is

$$R = R_0 + R_1 + R_2 +$$

$$= r_{12} + t_{12}r_{23}t_{21}e^{i2k_0nd}\left[1 + r_{21}r_{23}e^{i2k_0nd} + \left(r_{21}r_{23}e^{i2k_0nd}\right)^2 + \cdots\right]$$

$$= r_{12} + \frac{t_{12}r_{23}t_{21}e^{i2k_0nd}}{1 - r_{21}r_{23}e^{i2k_0nd}}$$

$$= \frac{1-n}{1+n} + \frac{4n}{(1+n)^2}\frac{(n-1.46)}{(n+1.46)}e^{i2k_0nd}\frac{1}{1 - \frac{(n-1)(n-1.46)}{(n+1)(n+1.46)}e^{i2k_0nd}} \tag{S27}$$

Here we used refractive indices of air and $SiO_2$ as 1 and 1.46, respectively.

The OPL was calculated by comparing the phase difference of the reflected light with and without the 2D material. Figure S8b shows the reference setup. Light is incident directly from air into infinite $SiO_2$ substrate. In this case the reflected amplitude is

$$R' = \frac{n_1 - n_3}{n_1 + n_3} \tag{S28}$$

So we get:

$$OPL = -\frac{\left(phase(R) - phase(R')\right)}{2\pi}\lambda \tag{S29}$$

where $\lambda$ is the wavelength of light. The magnitude of OPL is plotted in Figure 1f.

For phosphorene OPL calculations, we used the measured refractive index from bulk black phosphorus crystals ($n = 3.4$)[7].

## 5. Images and characterization of phosphorene flakes by PSI.

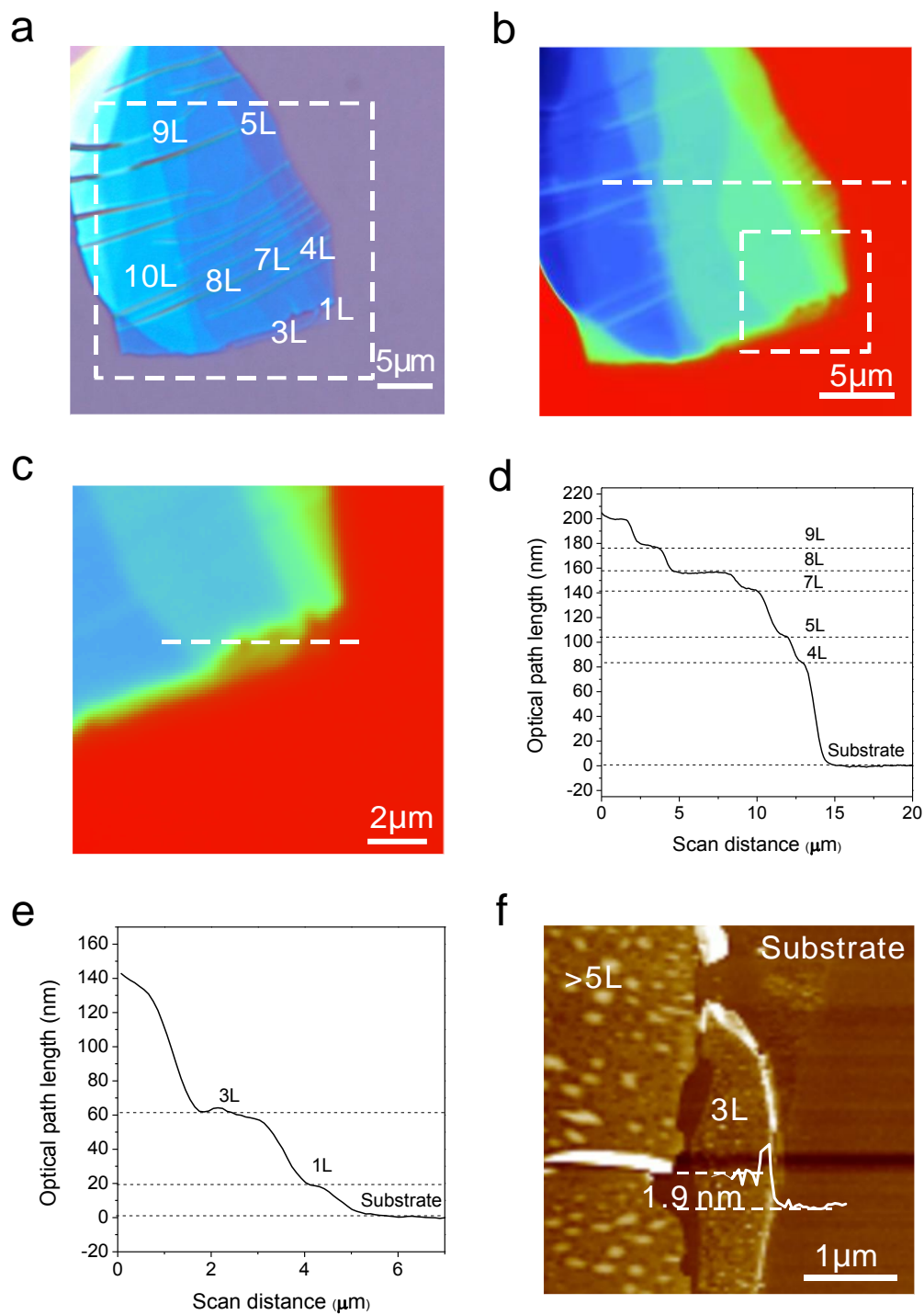

**Figure S9 | Images and characterization of exfoliated phosphorene. a,** Optical microscope image of a phosphorene flake containing multiple layers. **b,** PSI image of the phosphorene flake from the dash line box area indicated in (a). **c,** PSI image of the phosphorene flake from the dash line box area indicated in (b). **d** and **e** display the OPL measured by PSI versus position along the dash line in (b) and (c) respectively. **f,** AFM image of the 3L phosphorene flake.

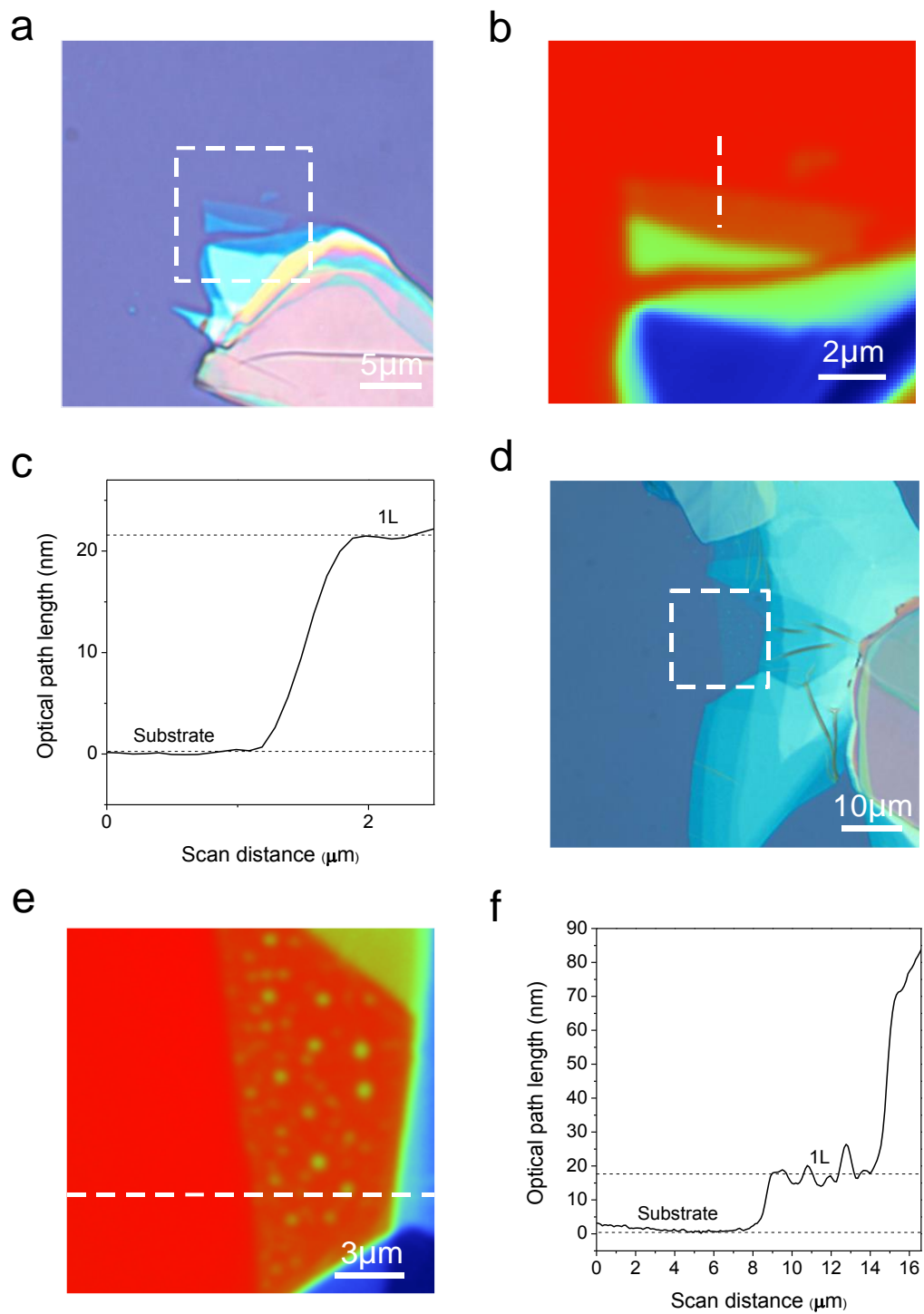

**Figure S10| Images of exfoliated 1L phosphorene flakes and their measured OPLs by PSI.**

**a,** Optical microscope image of a 1L phosphorene flake. **b,** PSI image of the 1L phosphorene flake from the dash line box area indicated in (a). **c**, OPL measured by PSI versus position along the dash line in (b). **d, e** and **f** indicate the optical microscope image, PSI image (from the dash line box in d) and OPL (along the dash line in e) of another 1L flake, respectively.

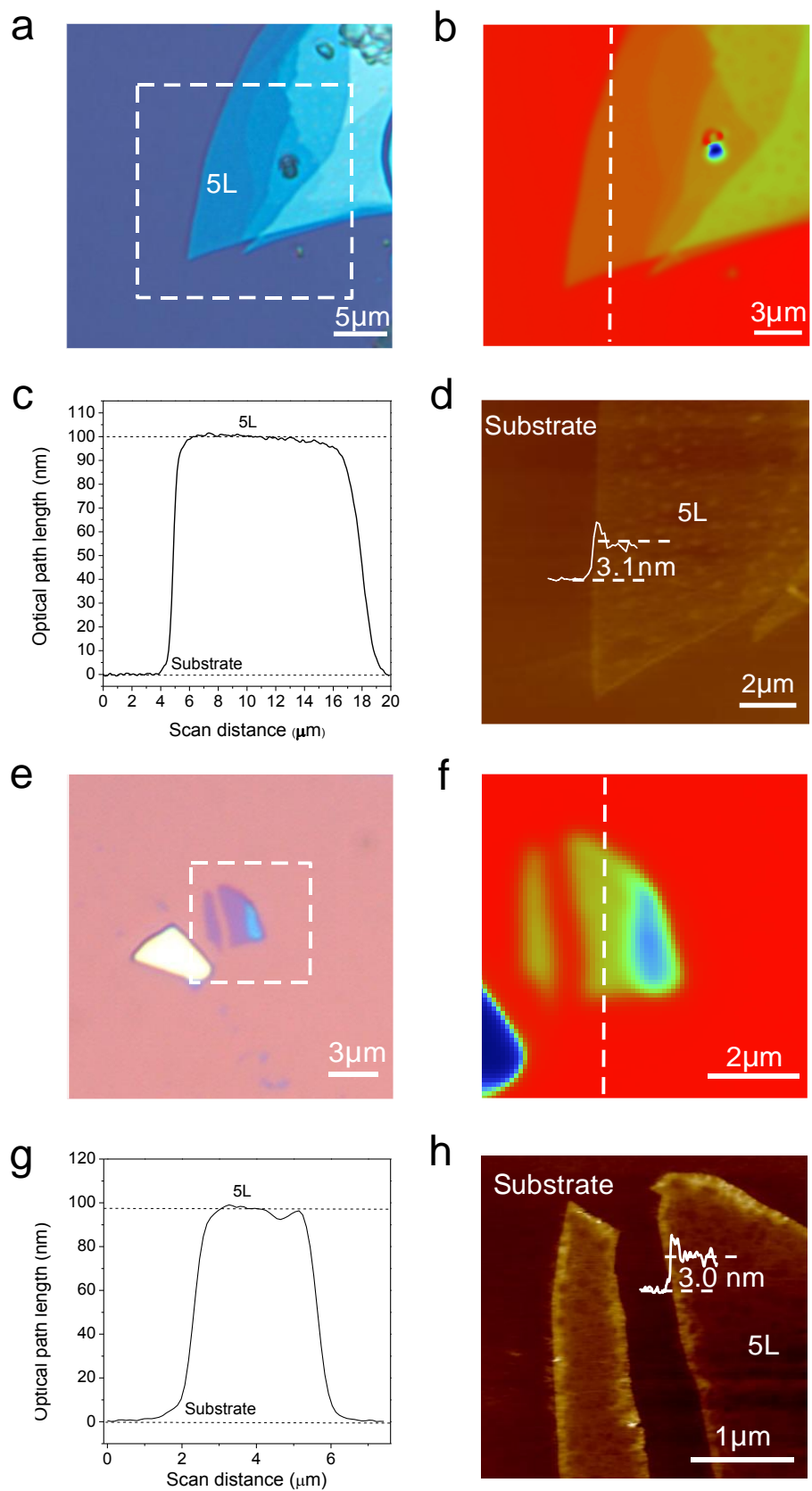

**Figure S11 | Images and characterization of exfoliated 5L phosphorene. a,** Optical microscope image of a 5L phosphorene flake. **b,** PSI image of the 5L phosphorene from the

dash line box area indicated in (a). **c,** OPL measured by PSI along the dash line in (b). **d,** AFM image of 5L phosphorene. **e, f, g** and **h** display optical microscope image, PSI image (from the dash line box area indicated in e), OPL (along the dash line in f) and AFM image of another 5L phosphorene flake, respectively.

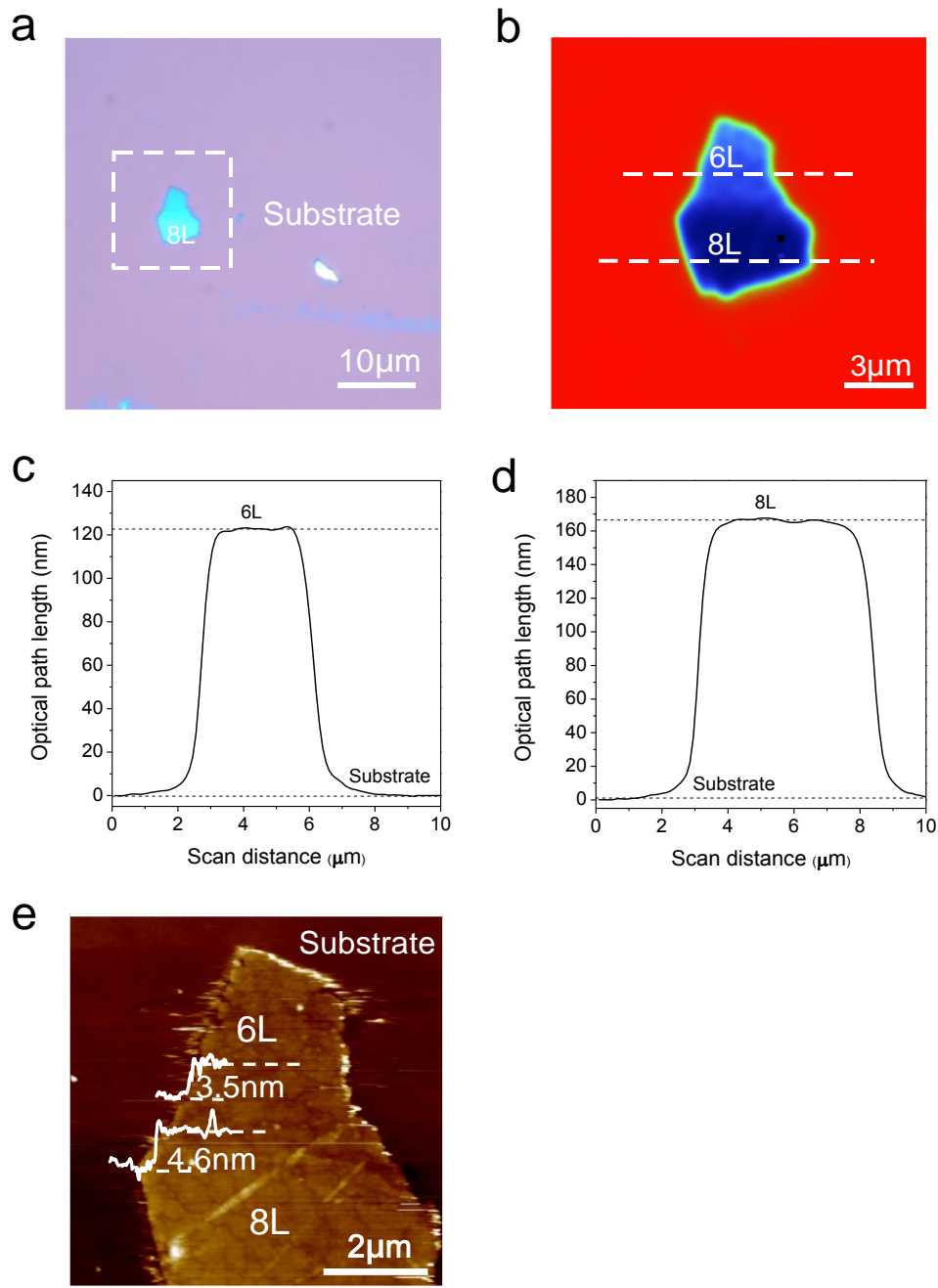

**Figure S12 | Images and characterization of exfoliated 6L and 8L phosphorene flakes. a,** Optical microscope image of a phosphorene flake containing 6L and 8L. **b,** PSI image of the

phosphorene flake from the dash line box area indicated in (a). **c** and **d** display OPL values measured by PSI from the 6L and 8L phosphorene along the dash lines in (b). **e,** AFM image of the 6L and 8L phosphorene.

**Reference**

1  Thilagam, A. Two-dimensional charged-exciton complexes. *Phys. Rev. B* **55**, 7804-7808 (1997).
2  Tran, V., Soklaski, R., Liang, Y. & Yang, L. Layer-controlled band gap and anisotropic excitons in few-layer black phosphorus. *Phys. Rev. B* **89**, 235319 (2014).
3  Qiao, J., Kong, X., Hu, Z.-X., Yang, F. & Ji, W. High-mobility transport anisotropy and linear dichroism in few-layer black phosphorus. *Nat. Commun.* **5** (2014).
4  de Groot, P. in *Optical Measurement of Surface Topography* (ed Richard Leach) Ch. 8, 167-186 (Springer Berlin Heidelberg, 2011).
5  Venkatachalam, D. K., Parkinson, P., Ruffell, S. & Elliman, R. G. Rapid, substrate-independent thickness determination of large area graphene layers. *Appl. Phys. Lett.* **99**, 234106 (2011).
6  Schreiber, H. & Bruning, J. H. in *Optical Shop Testing* 547-666 (John Wiley & Sons, Inc., 2006).
7  Nagahama, T., Kobayashi, M., Akahama, Y., Endo, S. & Narita, S.-i. Optical Determination of Dielectric Constant in Black Phosphorus. *J. Phys. Soc. Jpn.* **54**, 2096-2099 (1985).